\documentclass[12pt]{article}
\usepackage{amsmath,amssymb,graphicx,mathrsfs,hyperref,slashed}
\newcommand{\be}{\begin{equation}}
\newcommand{\ee}{\end{equation}}
\newcommand{\bea}{\begin{eqnarray}}
\newcommand{\eea}{\end{eqnarray}}

\def\({\left(} \def\){\right)}

\renewcommand{\baselinestretch}{1.5}
\begin{document}
\title{\vspace{-1.8in}
{Black holes as  collapsed polymers}}
\author{\large Ram Brustein${}^{(1)}$,  A.J.M. Medved${}^{(2,3)}$
\\
\vspace{-.5in} \hspace{-1.5in} \vbox{
 \begin{flushleft}
  $^{\textrm{\normalsize
(1)\ Department of Physics, Ben-Gurion University,
    Beer-Sheva 84105, Israel}}$
$^{\textrm{\normalsize (2)\ Department of Physics \& Electronics, Rhodes University,
  Grahamstown 6140, South Africa}}$
$^{\textrm{\normalsize (3)\ National Institute for Theoretical Physics (NITheP), Western Cape 7602,
South Africa}}$
\\ \small \hspace{1.07in}
    ramyb@bgu.ac.il,\  j.medved@ru.ac.za
\end{flushleft}
}}
\date{}
\maketitle
\begin{abstract}

We propose that a large Schwarzschild black hole (BH) is a bound state of highly excited, long,  closed strings at the Hagedorn temperature. The size of the bound state is smaller than the string random-walk scale and determined  dynamically by the string attractive interactions. It is further proposed that the effective free-energy density  of the bound state should be expressed as a function of its entropy density. For a macroscopic BH, the free-energy density  contains only  linear and  quadratic terms, in analogy with that of a collapsed polymer when expressed as a function of the polymer concentration. Using the effective free energy,  we derive scaling relations for the entropy, energy and size of the bound state and show that these agree with the scaling relations of the BH; in particular, with the area law for the  BH entropy. The area law originates from the inverse scaling of the effective temperature with the bound-state radius.  We also find that the energy density of the bound state is equal to its pressure.
\end{abstract}
\newpage
\renewcommand{\baselinestretch}{1.5}\normalsize

\section{Introduction}

Some recent discussions about  the information-loss paradox and  how  a
black hole (BH) radiates have led to an improved understanding of the constraints on the quantum state of a  BH. Given that  BHs  evolve in a unitary manner, the same as for any other quantum state, it
is by now clear that the once-standard Hawking model \cite{Hawk,info}  will have to undergo some rather dramatic modifications. This point has long been emphasized by Mathur \cite{Mathur1,Mathur2,Mathur3,Mathur4}, who then contends that  a much different entity  --- the  so-called fuzzball  model ---  can and should  effectively play the role of a BH. More recently, the ``firewall'' proponents, Almheiri {\em et al.} \cite{AMPS} (also see \cite{Sunny,Braun,MP}), have also realized  that the standard description of the BH interior  as an empty spacetime has to be substantially revised. A unitarily evolving BH cannot be empty as classical general relativity suggests.

The counter-proposals to Hawking's model---  whether a ball of fuzz or a wall of fire  --- are suggestive of  an excision of the BH interior itself; either as matter of principle in Mathur's setup  or as a matter of practice from the firewall perspective. That spacetime (effectively) ends at the BH horizon cannot be ruled out but it is also not the only logical possibility.  Indeed, we have proposed a model \cite{inny} that is premised on the existence of a specific  matter state for the  BH interior.

It was argued in \cite{inny} that the interior matter  must be in  a highly quantum state when described in terms of the Fock space of fields such as the spacetime metric. Here, highly quantum means that the density-matrix elements are small and, therefore, the quantum fluctuations are large and the state does not have a mean-field description in terms of the metric and the other fields.  This picture follows from the well-understood but often unappreciated fact that the emitted (Hawking) radiation is in a highly quantum state (see, {\em e.g.}, \cite{future}). Then, as the purifier of the BH radiation, the interior matter must similarly be in a highly quantum state. This argument becomes particularly acute when the age of the BH is parametrically similar to the Page time \cite{page}. The condition of a highly quantum state suggests  that, for instance, the ``leaky Bose-condensate''
description, as  advocated by Dvali and Gomez \cite{Dvali1,Dvali2,Dvali3}, cannot be valid. A condensate behaves semiclassically by design.

Whatever the state may be, the energy $E$ and entropy $N$ of the interior matter must account for the BH mass $M_{BH}$ and entropy $S_{BH}$, so that
$\;E=M_{BH}=\frac{R_S}{l_P^2}\;$ and $\;N=S_{BH}=\frac{R_S^2}{l_P^2}\;$.~\footnote{Conventions: $R_S$ is the Schwarzschild radius, $l_P$ is the Planck length, we work in units in which $c$, $\hbar$, $k_B$ are set to unity, the spacetime dimensionality is four for concreteness and numerical factors of order unity are typically disregarded. Also, $l_s$ is the string length, which is
often set equal to one, and $g_s$ is the string coupling.}

But a highly quantum state further requires a large number ($N$) of single-occupancy energy levels. For the BH interior, this means a parametrically  larger number of available levels as each occupied one should have a net energy  of order $1/R_S$ to account for the BH mass. Such a large density of states is unusual in the context of standard  quantum field theory. Yet, this must be the case for the state of the BH interior if it is meant to be the purifier of the external radiation.

We also argued in \cite{inny} for the ``no-remnant'' condition. This means insisting that the Bekenstein--Hawking area law \cite{Bek} be preserved throughout the interior and not just at the horizon.  In other words, the entropy  that is enclosed by a spherical shell at {\em any}  radius $\;r< R_S\;$ cannot exceed the area of this shell in Planck units.   One can notice that this  condition immediately rules out any scenario which remotely resembles the general-relativistic picture of mostly empty space with a
highly dense core. One rather finds that the energy and entropy density
go respectively as
$\;\rho=  1/(r^2\; l_P^2)\;,
$
and
$\;
n= 1/(r\; l_P^2)\; ;
$
so that, in Planck units,
$\;
n= \sqrt{\rho}\; .
$
This is the equation of state of a thermal gas in
$(1+1)$-dimensions, for which the pressure $p$ and energy density are equal,
$\;
p= \rho\; .
$
Other authors have conjectured that the same equation of state could be relevant to the BH interior \cite{Page2,Hooft}.

There are two relevant points regarding the results of this model. First, as already mentioned in \cite{inny}, the relation~$p=\rho$ coincides with the saturation of the causal entropy bound \cite{CEB}, which should then signify the breakdown of semiclassical physics \cite{maxme}. Such a breakdown is consistent with the notion of a highly quantum state. Second, this same  equation of state turns up for only one
type of {\em fundamental} matter theory; namely, a theory of highly excited, long, closed strings at temperatures close to the Hagedorn temperature \cite{AW}. This means that, whatever fundamental description underlies the interior matter, it should have an effective description  in terms of string theory. In this paper, we are proposing  that the interior of a  BH is indeed composed mainly of highly excited, closed strings.

In what follows, we will first recall in Section~2 the statistical mechanics of  highly excited, closed strings in a bounded volume, following a discussion by Lowe and Thorlacius \cite{LT}. Then, in Section~3, we will consider a dynamically formed bound state of highly excited strings and its relation to the collapsed polymer and, from this, obtain scaling relations among the energy, entropy and size of the bound state. Additionally, we will show how the equation of state $p=\rho$ arises. In Section~4, a dictionary will be  set up between BH geometric quantities and the stringy bound-state parameters, and it will be shown that, as a consequence of the polymer scaling relations, the area law for BH entropy emerges. We conclude in Section~5.

Before moving on, let us clarify as to what our proposal is {\em not}. Although there are some superficial similarities, the proposed model is not that of the well-known string--BH correspondence \cite{corr1,corr2}. There are two significant differences. First,  this correspondence advocates that, as the string or gravitational coupling is dialed up, a long, closed string will eventually undergo a type of phase transition that transforms it into a BH. This is supposed to happen just as  the BH  radius  reaches the string scale. Then, once this correspondence point is passed, the string is replaced by a BH. (See, however, \cite{GVtalk} where the possibility of a stringy composition of larger BHs was considered.) Conversely, we are suggesting that the interior of even a macroscopically large BH has a stringy composition, regardless of what type of matter collapsed to form it.

Second, it is either implicit or explicit throughout the correspondence literature that the coupling can, in principle, be dialed up as high as one would like, well past the correspondence point. On the contrary, a central part of our proposal is that the string  coupling for a BH is given by a certain critical value that scales with the inverse of the entropy, $\;g^2_c \propto 1/N\;$, but with  $\;g^2_c N\gg 1\;$ and yet $\;g_c^2\ll 1\;$. This is distinct from the relation $\;g^2_c N= 1\;$  that has long been advocated by Dvali and Gomez \cite{Dvali1,Dvali2,Dvali3} as well as by the current authors \cite{RB,flucyou}.

\section{Review of highly excited strings in a bounded region}

To gain some understanding about the basic nature of a bound state  of strings, we will appeal to an old but highly relevant paper by Lowe and Thorlacius (LT) \cite{LT}. This paper expresses  the results of an even older paper by Salomonson and Skagerstam (SS) \cite{SS} in a way that  pertains to the problem at hand. The findings of both \cite{SS} and \cite{LT} are based on two properties of interacting, long strings: (1) That the total length is conserved due to the conservation of energy and (2) that the probability for interaction is proportional to the length of the string.

The SS--LT model was later on supplemented by a more detailed discussion in terms of the ``thermal-scalar" formalism in \cite{HP} and then directly in terms of  the world-sheet description of (open) bosonic strings \cite{DV}. The latter was extended to a variety of closed-string theories in \cite{chialva}. These results clarify that the somewhat heuristic discussion of \cite{LT} is indeed valid.

We would like to use the LT description to motivate our upcoming proposal for an effective free energy for the interacting strings. These authors considered a high-temperature collection of interacting, closed strings in a finite volume such that the total length of string far exceeds the spatial dimensions of the ``box''. They consider the box to be external to the strings; for example, strings moving on a toroidal space of a give size. As such, LT  assume a total interaction energy that is much smaller than
the string tension, and  so they could neglect four and higher-order string interactions.

The distinction between their scenario and ours is that, for us,  the  finite volume is due to  the self-binding of the strings. Hence, in our case, the interaction energy  is parametrically the same as the string tension. However, as
 will be seen later, the four and higher-order string interactions can be neglected in our case as well.

We will, for the sake of completeness, briefly sketch   the pertinent analysis of \cite{LT}. In particular, LT discuss the Boltzmann equation for long, closed strings when only three-string interactions are relevant. In  $\;l_s=1\;$ units, this is
\bea
 {\dot n}(\ell) \;=\; k \frac{g_s^2}{V}\Big[&-&\frac{1}{2}\ell^2 n(\ell)-\int^{\infty}_0 d\ell'\;\ell'n(\ell')
\ell n(\ell)
 \nonumber \\
&+&\frac{1}{2}\int^{\ell}_{0} d\ell'\;\ell'(\ell-\ell')n(\ell')n(\ell-\ell')
+ \int^{\infty}_{\ell} d\ell'\; \ell' n(\ell')\;\Big]\;,
\label{Boltz}
\eea
where $n(\ell)$ is the average number of strings of length $\ell$ (not to be confused with the entropy density $n$), the dot denotes a time derivative,  $k$ is a numerical factor and each of  the four terms on the right represents a different type of three-string interaction. The first term  represents the possibility that a string of length $\ell$ splits into two shorter ones, while the second term accounts for  two strings of lengths $\ell$ and $\ell'$ joining together to form one of length $\ell+\ell'$. The third term results from two shorter strings joining to make a longer string of length $\ell$ and, finally,  the last term accounts for a longer loop splitting into two shorter ones, one of which is of length $\ell$. LT disregard
higher-order interactions by assuming  that
the string coupling is sufficiently small.

The equilibrium condition $\;{\dot n}(\ell)=0\;$  leads to the distribution
\be
n(\ell) \;=\; \frac{1}{\ell} e^{-{\ell}/{L}}\;,
\label{above}
\ee
where $\;L=\int_0^{\infty}d\ell\;\ell n(\ell)\;$.
This outcome amounts to a total number of approximately $\ln{L}$ strings whose total length is $L$ to an exponentially good approximation.

Let us define $N$ as the number of ``string bits" of length $l_s$, so that $\;N=L/l_s\;$.  The string entropy is then essentially that  of an $N$-step random walk,
which is the logarithm of the number of different walks. If, for example, each step is $\pm 1$, then $\;S=\ln{2^N}\sim N\;$.

The strings are distributed such that, on average, there is one string at each length interval from
$\;\ell=\ell_0\;$  to $\;\ell=e \ell_0\;$ up to $\;\ell=L\;$.  The  distribution of lengths is flat ---  the probability of finding any one bit on  a string of length $\ell$ does not depend on $\ell$, as $\;\ell n(\ell)\;$ is approximately  constant. Hence, most of the bits will be found on the  long strings, which  will then make the dominant contribution to the entropy and energy,  both of which  scale with the total length of string.

\section{A bound state of highly excited strings: The collapsed polymer}

Our current objective is to describe the BH as a bound state of highly excited strings. In contrast to the previous discussion, where the bounded region was enforced on the strings by putting them in an external ``box", we would like to consider a scenario where strong interactions lead to the formation of a bound state of strings, the volume of which is determined dynamically.

When the strings are free and put in a large-enough volume, the linear size $R$ of the region that is occupied by a string
of length $\;N=L/l_s\;$  is the random-walk scale, $\;R\sim \sqrt{N}\;$. If the string interactions were repulsive, then $R$ would increase and  be parametrically larger than $\sqrt{N}$. Conversely, attractive interactions would
lead to a parametrically decreasing value of $R$  \cite{HP,DV}. In general, we expect $\;R\sim N^\nu\;$ for some
$\nu$. Clearly, the nature and strength of the string interactions determines the properties of the bound state.
It  will be shown that, to match the properties of a macroscopic BH, these interactions have to be sufficiently strong.

The current situation is analogous to the Flory--Huggins theory of polymers \cite{flory}. (See also the books by De Gennes \cite{bookdg} and Doi and Edwards \cite{bookde}.) This theory is reconsidered in \cite{degennes} and reviewed, for example, in \cite{polymer}. The transition temperature when the polymers become effectively tensionless is known as the Flory temperature and $\nu$ is called the Flory exponent. The linear size $R$ is
known as the gyration radius of the polymer. When the polymer interactions are repulsive, the effects are similar to an excluded-volume random walk. In  the polymer literature, the case of attractive interactions  is  referred to as ``negative  excluded volume". In this case, the gyration radius of the polymer is smaller than $\sqrt{N}$ and it  is then called a ``collapsed polymer".

What we would like to find --- an effective description of a bound state of highly excited strings --- is therefore already provided by the theory of collapsed polymers, with the appropriate adaptations. Let us then recall the Flory effective free energy for the case of the collapsed polymer (see, {\em e.g.}, \cite{polymer}),
\be
-\left(\frac{F}{VT}\right)_{\rm CP}\;=\;c - \frac{1}{2} \upsilon c^2  +{\cal O}[c^3]\;.
\label{cp}
\ee
Here, $c$  is the entropy density (monomer concentration)  and $\upsilon$
is the absolute value of the negative excluded volume. The relevance of the higher-order terms is controlled by the value of $\upsilon$. If the negative excluded volume is parametrically smaller than the volume that is occupied by the polymer, then the tertiary and higher-order interaction terms are not important, just as in the previous LT discussion.

Guided by Eq.~(\ref{cp}), we can write an effective free energy for highly excited strings (near the Hagedorn temperature $\;T_{Hag}\sim 1$) in terms of their
presumptive entropy density $\;c=N/V\;$:
\be
-\left(\frac{F}{V T_{Hag}}\right)_{strings}\;= \;\epsilon c - \frac{1}{2} \upsilon c^2 +{\cal O}[c^3]\;
\label{FES}
\ee
or, equivalently,
\be
-\left(\frac{F}{T_{Hag}}\right)_{strings}\;=\; \epsilon N - \frac{1}{2} \frac{\upsilon}{V} N^2 +{\cal O}\left[\frac{N^3}{V^2}\right]\;.
\label{FES1}
\ee
To proceed, what is then required is an explicit form for  the excluded volume
$\upsilon$ and for the additional (dimensionless) parameter $\epsilon$ in the case of the highly excited strings.

The parameter $\epsilon$ will help us to distinguish between two notions of energy or temperature. The energy of  free strings goes as
$\;
E_{free}=N\;
$
and, correspondingly, the temperature is  close to the Hagedorn temperature ({\em i.e.}, one in string units).
However, the strings are strongly bound, and so the
net  energy of the  bounded string state is much smaller than the energy of free strings:
$\;
E_{bound}= E_{free} -E_{int}\ll E_{free}\;,
$
where $E_{int}$ indicates the interaction energy. The parameter $\epsilon$ then allows for the reduction $\;E_{bound}=N\epsilon\;$ and therefore $\;\epsilon \ll 1\;$. The parameter $\epsilon$ can also be viewed as the relative reduction of the temperature from the Hagedorn value to some much lower value (which, not surprisingly, will turn out to be the Hawking temperature).

In the polymer literature, the excluded volume $\upsilon$ is typically regarded as a phenomenological parameter. Since, in our case, it can be attributed to the effects of local string interactions, it is identified  as $\;\upsilon = g_s^2 l_s^3=g_s^2\;$.

The effective free energy is then
\be
-\left(\frac{F}{V T_{Hag}}\right)_{strings}\;= \;\epsilon c - \frac{1}{2} g_s^2  c^2 +{\cal O}[c^3]\;
\label{FES2}
\ee
or
\be
-\left(\frac{F}{ T_{Hag}}\right)_{strings}\;= \;\epsilon N - \frac{1}{2} g_s^2  \frac{N^2}{V} +{\cal O}\left[
\frac{N^3}{V^2}\right]\;,
\label{FES3}
\ee
from which  $\epsilon$  can indeed be identified
as the effective (dimensionless) temperature, $\;T=\epsilon\;$.

It is now a matter of evaluating the entropy density, energy density and pressure via the standard thermodynamic definitions and then substituting  for $c$ via the equilibrium condition $\;c=\epsilon/g_s^2\;$,
which follows from
$\;\frac{\partial F}{\partial c}=0\;$.
Similarly for $N$,
\be
\label{consol}
\frac{N}{V} \;=\; \frac{\epsilon}{g^2_s}\;.
\ee

What  one finds is that
\bea
n &=& -\left.\frac{1}{V}\frac{\partial F}{\partial \epsilon}\right|_{c=\epsilon/g_s^2}\;=\; \frac{\epsilon}{g_s^2}\;,
\nonumber \\
\rho &=& \left[\frac{F}{V} + \epsilon s\right]_{c=\epsilon/g^2_s} \;=\; \frac{1}{2}\frac{\epsilon^2}{g^2_s}\;,
\nonumber \\
p&=& -\left.\frac{\partial F}{\partial V}\right|_{c=\epsilon/g^2_s} \;=\; \frac{1}{2}\frac{\epsilon^2}{g^2_s}\;.
\eea
In addition, one can calculate the effective tension and confirm that it vanishes at the Hagedorn (or Flory) temperature as expected,
\be
\sigma\;=\; \left.\frac{\partial F}{\partial L}\right|_{c=\epsilon/g^2_s} \;=\;\left.\frac{\partial F}{\partial N}\right|_{c=\epsilon/g^2_s} \;=\;0\;.
\ee

Not only do we obtain the  anticipated relations for $n$ and $\rho$  ({\em cf}, Section~1) but also the signature equation of state  for  high-temperature, tensionless strings, $\;p=\rho\;$.

Let us now determine the scaling relation for the effective temperature. Here, we use the fact that, at long distances, the string interactions are dominated by gravity \cite{HP,DV}. Hence, the total interaction energy of the bound state of strings is parametrically equal  to its total gravitational energy,
\be
g^2_s \frac{N^2}{V} \;=\; G_N \frac{E^2}{R}\;,
\label{grav-energy}
\ee
where the left-side follows from an inspection
of the LT Boltzmann Equation (\ref{Boltz}) and
the right-hand side is the Newtonian potential with
$\;G_N=l_P^2=g^2_s l_s^2 = g^2_s\;$.
The energy $E$ on the right-hand-side of Eq.~(\ref{grav-energy}) must be the bound-state energy $\;E_{bound}=\epsilon N\;$, and so
\be
\frac{R}{\epsilon^2 V}\;=\; 1\;.
\ee
Then, since $\;V=R^3\;$, it follows that  $\;\epsilon=1/R\;$
({\em i.e.}, the Hawking temperature).

The self-consistency of this framework necessitates
the following set of hierarchies:
$\;\epsilon\ll 1\;$, $\;g^2_s \ll 1\;$, $\;\epsilon/g^2_s \ll 1\;$ and $\;g^2_s N \gg 1\;$.
These guarantee that the higher-order string interactions coming from either $\alpha^\prime$ corrections, loop corrections or their combination are indeed suppressed.

The last claim can be substantiated as follows: The relative strength of  $(n+2)$-string interactions is determined
by $g_s^{2n}$,  along with a combinatoric  enhancement factor $N^n$ and a  volume suppression factor $V^{-n}$.
In string units, the multi-string  interaction strength then goes as
\be
\lambda_{n+2}\;\sim\;  \left(\frac{g_s^{2}N}{V}\right)^n\;=\; \epsilon^{n}\;,
\ee
where we have used Eq.~(\ref{consol}). One can see that these are $\alpha^\prime$ corrections,
being proportional to powers of $l_s/R$.  Similarly, higher-order string-loop corrections and combinations thereof are suppressed.

Now, as an added bonus, one can see how the area scaling for the entropy comes about. Since $\;N=c V = \epsilon V/g^2_s\;$, it follows that $\;N= R^2/g^2_s\;$, an outcome
that is readily generalized to higher dimensions.
This area scaling comes about dynamically because the effective temperature of the bound state scales as $1/R$. We will show shortly that these scaling relations amount, not only to fixing the area scaling of the entropy, but also to (almost) fixing the coefficient.

\section{The string--black hole dictionary}

At this point in the discussion, it is useful to set
up a dictionary that translates statements about the bound state of
strings into statements  about the BH geometry.
It should be kept in mind that $\;G_N=g^2_s\;$
and $\epsilon=1/R\;$ in string
units.

Clearly, the scaling relations of the BH mass and entropy match those
of the bound state since  $\;E\sim \epsilon^2 V/g^2_s = R/G_N\;$ and $\;N \sim \epsilon  V/g^2_s = R^2/G_N\;$. The scaling relations can, however, only fix the coefficient of the product of the BH entropy and
Hawking temperature,
\be
S_{BH} T_H \;=\; \frac{R_S}{4 G_N}\;.
\label{action}
\ee
This is a dimensional quantity with units of action, which in our units is equivalent to inverse length.

When comparing the bound state to the BH, we have to compare them for the same fixed mass $\;E_{bound}=M_{BH}\;$ and for the same choice of radius $\;R=R_S\;$,  as these are the observables
which can be accessed by an outside observer.  For the bound state, the scaling relations  amount to
\be
N \;\sim\; E_{bound}R \;=\; C_1 \frac{R^2}{g^2_s}\;,
\ee
where $C_1$ is a  numerical constant that can depend on some other numerical constants in the effective action,  the
precise   relation between $\epsilon$ and $R$, {\em etc}. Similarly, the scaling relations determine that the temperature  satisfies
\be
T\;=\; C_2 \frac{1}{R}\;.
\ee

It follows that, for the bound state,
\be
N T\; =\; C_1 C_2 \frac{R}{g^2_s} \;=\; C_1 C_2 \frac{R}{G_N}\;.
\ee
Since this quantity has units of $1/l_s$, fixing the the relation to yield  $\;C_1 C_2=1/4\;$ is  no different that choosing the units so that the coefficients are absorbed into a redefinition of the string length.

To do better, we have to calculate the entropy of the bound state microscopically; for instance, by seeing how
interactions modify  the random-walk picture.

\section{Conclusion}

We have proposed that the interior of a semiclassical  BH is composed of a highly excited state of closed strings that is dominated by a few long strings. The motivation for this proposal follows from two simple but powerful ideas: The state of the BH must be highly quantum to act as the purifier of the Hawking radiation and no closed surface can contain more than an area-law's worth of entropy.

We have shown that the scaling properties of the stringy bound state match those of the BH and is able to correctly account for BH thermodynamics in a self-consistent manner. The collapsed-polymer picture agrees with all the pieces of evidence that have been gathered so far about BHs, and we are unable to identify an argument saying that such a state cannot describe a macroscopic BH.

If the BH interior is indeed a bound state of strings, it opens up the door to some exciting possibilities. One can now ask,  for instance, about   the plight of an object falling into a BH (and finally expect a rigorous answer) or find out whether there is some observable effect of the inner-structure of the BH when two BHs collide \cite{ligo}. Also, is there a deeper meaning to the state-of-the-art  calculations of (near) extremal BH entropy \cite{SV}? One potential obstacle is the breakdown of  mean-field techniques when applied to  such  highly quantum systems. On the other hand, the analogy with the collapsed-polymer model suggests that there is  already an existent  field of research that is geared toward circumventing this same hurdle.

In a forthcoming paper we will discuss the dynamical properties of the bound state.

\section*{Acknowledgments}

We would like to thank Itzhak Fouxon, Rony Granek, Sunny Itzhaki, Oleg Krichevsky, Costas Skenderis, Marika Taylor and Gabriele Veneziano  for valuable discussions.
The research of AJMM received support from an NRF Incentive Funding Grant 85353, an NRF Competitive Programme Grant 93595 and Rhodes Research Discretionary Grants. AJMM thanks Ben Gurion University for their  hospitality during his visit.

\end{document}